\documentclass[aps,prl,twocolumn,superscriptaddress,floatfix,nofootinbib]{revtex4-2}
\usepackage{epsfig,amsmath,amssymb,color,comment,physics}
\usepackage[makeroom]{cancel}
\usepackage[caption=false]{subfig}
\usepackage{mathrsfs}
\usepackage[countmax]{subfloat}
\usepackage[normalem]{ulem}
\usepackage[english]{babel}
\usepackage{dsfont}
\usepackage{float}
\usepackage[bookmarks=true,colorlinks,linkcolor=black,urlcolor=NavyBlue,citecolor=RoyalBlue]{hyperref}
\usepackage[dvipsnames]{xcolor}
\usepackage{soul} 
\usepackage{color, xcolor}

\hypersetup{
    unicode=false,     
    pdftoolbar=false,  
    pdfmenubar=true,   
    pdffitwindow=false, 
    pdfstartview={FitH},
    pdftitle={},    
    pdfauthor={Authors},     
    pdfsubject={},   
    pdfcreator={},   
    pdfproducer={}, 
    pdfnewwindow=true,
    colorlinks=true,
    linkcolor=black,
    citecolor=blue, 
    filecolor=magenta,
    urlcolor=blue
}

\usepackage{amsfonts}

\makeatother

\begin{document}

\title{Low-Energy Free-Electron Nonclassical Lasing}

\author{Mai Zhang}
\affiliation{Laboratory of Quantum Information, University of Science and Technology of China, Hefei, Anhui 230026, China}
\affiliation{CAS Center For Excellence in Quantum Information and Quantum Physics, University of Science and Technology of China, Hefei, Anhui 230088, China}
\affiliation{Anhui Province Key Laboratory of Quantum Network, University of Science and Technology of China, Hefei, Anhui 230088, China}

\author{Yu Wang}
\affiliation{Laboratory of Quantum Information, University of Science and Technology of China, Hefei, Anhui 230026, China}
\affiliation{CAS Center For Excellence in Quantum Information and Quantum Physics, University of Science and Technology of China, Hefei, Anhui 230088, China}
\affiliation{Anhui Province Key Laboratory of Quantum Network, University of Science and Technology of China, Hefei, Anhui 230088, China}

\author{Chang-Ling Zou}
\affiliation{Laboratory of Quantum Information, University of Science and Technology of China, Hefei, Anhui 230026, China}
\affiliation{CAS Center For Excellence in Quantum Information and Quantum Physics, University of Science and Technology of China, Hefei, Anhui 230088, China}
\affiliation{Anhui Province Key Laboratory of Quantum Network, University of Science and Technology of China, Hefei, Anhui 230088, China}

\author{Lei Ying} \email{leiying@zju.edu.cn}
\affiliation{School of Physics, Zhejiang Key Laboratory of Micro-nano Quantum Chips and Quantum Control, Zhejiang University, Hangzhou 310027, China}

\author{Qiongyi He}
\affiliation{State Key Laboratory for Mesoscopic Physics, School of Physics and Frontiers Science Center for Nano-Optoelectronics, Peking University, Beijing 100871, China}

\author{Guang-Can Guo}
\affiliation{Laboratory of Quantum Information, University of Science and Technology of China, Hefei, Anhui 230026, China}
\affiliation{CAS Center For Excellence in Quantum Information and Quantum Physics, University of Science and Technology of China, Hefei, Anhui 230088, China}
\affiliation{Anhui Province Key Laboratory of Quantum Network, University of Science and Technology of China, Hefei, Anhui 230088, China}

\author{Chun-Hua Dong}
\email{chunhua@ustc.edu.cn}
\affiliation{Laboratory of Quantum Information, University of Science and Technology of China, Hefei, Anhui 230026, China}
\affiliation{CAS Center For Excellence in Quantum Information and Quantum Physics, University of Science and Technology of China, Hefei, Anhui 230088, China}
\affiliation{Anhui Province Key Laboratory of Quantum Network, University of Science and Technology of China, Hefei, Anhui 230088, China}

\begin{abstract}
Harnessing a beam of slow free electrons in artificial photonic structures offers a powerful, tunable platform for generating nonclassical light without the need for heavy physical equipment. Here we present a theory of nonclassical lasing, demonstrating how incoherent electrons in photonic crystal cavities can coherently emit photons through collective dynamics. When photon emission rate exceeds cavity losses, nonclassical lasing with sub-Poissonian photon statistics emerges, driven by multi-photon Rabi oscillations. At specific coupling strengths, quantum state trapping effect emerges, producing high-fidelity Fock states at room temperature (e.g. nearly 90\%-fidelity of four-photon Fock state). Notably, the frequency of the emitted photons can be readily tuned via the velocity of the injected electrons to match cavity modes.  This approach supports photonic integration and offers a scalable, energy-efficient platform for room-temperature quantum light sources and advanced studies in quantum electrodynamics.
\end{abstract}

\maketitle

{\it Introduction}---Generation of various quantum states of light~\cite{caspani2017integrated,kolobov1999spatial,lewenstein2021generation,shields2007semiconductor,sun2023catsate,sychev2017enlargement} facilitates the development of novel technologies across quantum communication~\cite{ralph1998teleportation,gisin2007quantum,lo2014secure,armstrong2015multipartite,xu2020secure}, quantum metrology~\cite{giovannetti2011advances,aasi2013enhanced}, and other emerging fields~\cite{slussarenko2019photonic,zhong2020quantum,le2018remote,neergaard2006generation,ourjoumtsev2006generating,vlastakis2013deterministically}.
Conventionally, its creation is primarily accomplished by exciting real quantum systems---specifically bound-particle systems exhibiting discrete energy levels like atoms and quantum dots---to induce the emission of Fock states~\cite{hijlkema2007single,senellart2017high,morie2025wang}, entangled states~\cite{entangled2006akopian}, and other quantum light.
Alternatively, free electrons, unlike bound-particle systems,  possess a continuous energy spectrum~\cite{barwick2009photon,ruimy2025free,velasco2025free}. 
This enables coupling with photonic modes across a wide frequency range, thereby facilitating the generation of light at arbitrary wavelengths~\cite{korbly2005observation,adamo2009light,roques2019towards,shentcis2020tunable,karnieli2022cylindrical}.
For example, when a beam of free electrons is accelerated to relativistic speeds and passed through a periodic magnetic field, it can emit a large number of photons with a broad frequency spectrum, known as the synchrotron radiation~\cite{o2001free,emma2010first,oepts1995free,deacon1977first,Kling2017Quantum}.
More recently, experiments employing lower-energy free electrons have demonstrated strong coherent photon-electron interactions within optical cavities~\cite{wang2020coherent,kfir2020controlling,dahan2021imprinting,henke2021integrated,feist2022cavity,Arend2025Electrons}. 
Concurrently, theoretical work proposed that low-energy electrons, due to their nonlinear dispersion relation, realize the Jaynes–Cummings model, potentially enabling efficient generation of single photons and photon pairs~\cite{fan2023JC,karnieli2021superradiance}.

These above progress on strong photon-electron coherent interactions comprises the low-energy free-electron system as a promising platform for frequency-agile quantum light sources.
However, prior studies, confined to single-electron schemes~\cite{ben2021shaping,bendana2011single,fan2023JC,karnieli2021superradiance,sun2023catsate,Arend2025Electrons}, are limited by low emission rates constrained by the inherent electron-photon coupling strength~\cite{Xie2025Maximal,Zhao2025Upper}, thus impeding applications demanding high-brightness quantum light.
Inspired by micro-maser in a stream of excited atoms that interacts with an optical cavity to produce coherent radiation~\cite{rempe1987observation,filipowicz1986theory,rempe1990observation,weidinger1999trapping}, the system with many incoherent slow-flying electrons offers a practical pathway to amplify quantum light through collective effects.

In this paper, we present a comprehensive theoretical framework for low-energy free-electron-driven lasing process by elucidating multi-electron dynamics in photonic cavities. 
Unlike conventional platforms that require coherence between electrons, low-energy free-electron beams inherently function as tunable gain media. Despite their intrinsic incoherence, they can still emit photons coherently into multiple cavity modes when their momenta satisfy the phase-matching conditions.
Notably, the deterministic electron-photon coupling phases can be tuned by the electron-photon interaction time and the structure of photonic crystal (PhC) cavity. These can suppress phase randomization inherent in classical lasers. 
By tuning the coupling strength and interaction time, this system can generate both nonclassical laser and high-purity Fock states, even achieving a fidelity of for four-photon Fock states exceeding $90\%$ at room temperature. Our findings suggest that the proposed system is a feasible and scalable platform for room-temperature quantum light sources.

\begin{figure}[t]
\centering
\includegraphics[clip,width=1\columnwidth]{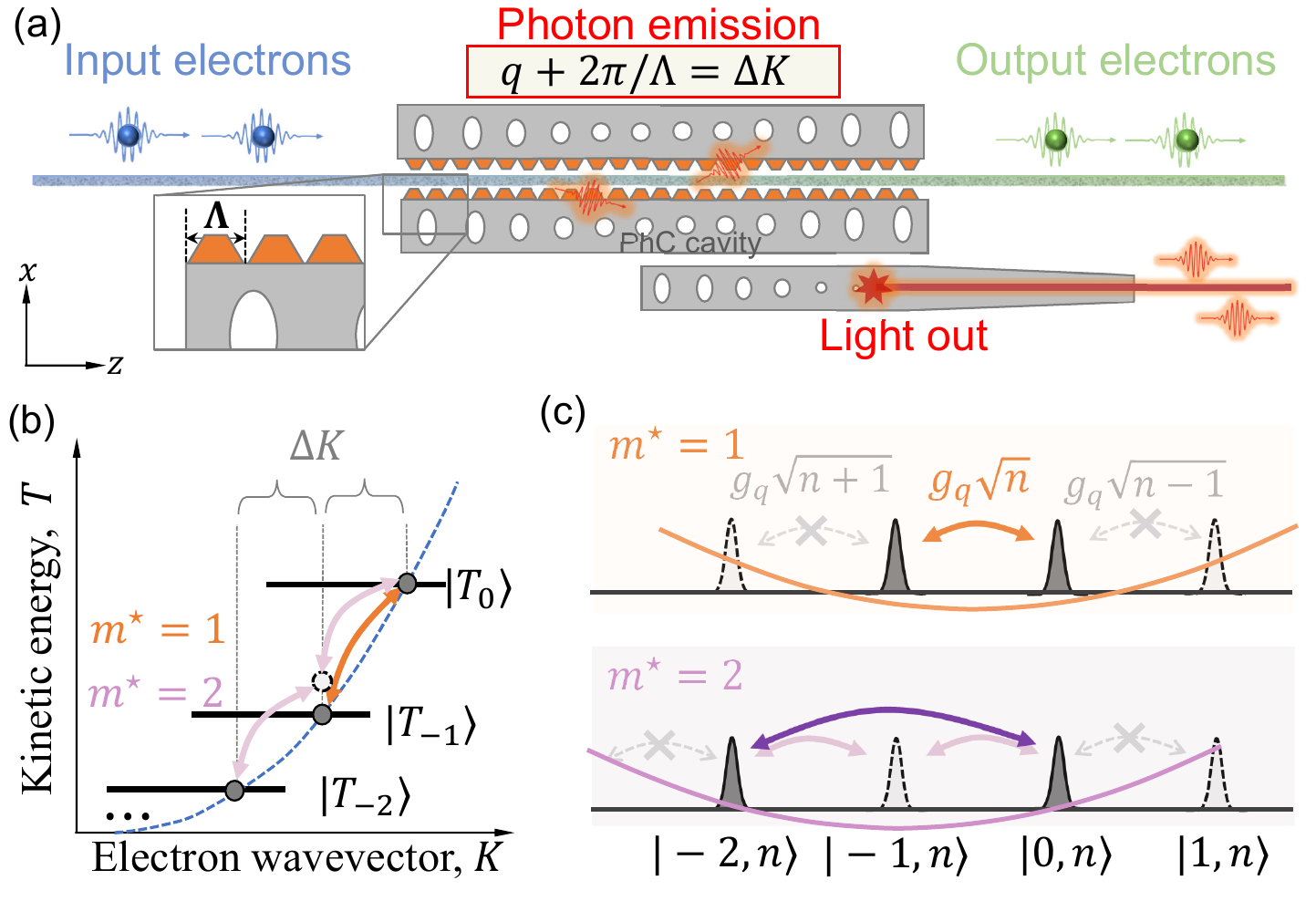}
\caption{(a) Sketch of the lasing process. Electrons are injected and interact with the optical mode in the FP PhC cavity with structures periodicity $\Lambda$. (b) Energy levels of a low-energy free electron when emitting the photons with the recoil of $\Delta K$. Orange arrow indicates single-photon resonance process, while purple arrows represent two-photon process. (c) Effective formulation of the electron-photon coupling dynamics for single-photon (upper) and two-photon (lower) resonance conditions.}\label{fig:1}
\end{figure}

{\it System and model}---As illustrated in Fig. \ref{fig:1}(a),  we consider a beam of low-energy electrons
injected into a PhC cavity. The periodicity of the cavity $\Lambda$ enables phase-matching condition as $\Delta K-q-{2\pi}/{\Lambda}$=$0$, where $\Delta K$ and $q$ denote the recoil wavevector of electrons and photon wavevector, respectively. The momentum bandwidth of the interaction is negligible to $\hbar\Delta K$, justifying the effective model presented below~\cite{kfir2019entanglements}. Then, the electron momenta are $Q_{m}=Q_{0}+m\hbar\Delta K$ with $m\in \mathbb{Z}$.  The system wavefunction is $|\varphi(t)\rangle=\sum_{n=0}^{\infty}\sum_{m=-\infty}^{n}c_{m,n}(t)|m,n\rangle$, where $|m,n\rangle=|Q_m\rangle\otimes|n-m\rangle$ and  `$m,n$' label number of absorbed photons and initial photon number, respectively. Here, the initial state is $|\varphi(0)\rangle=|Q_{0}\rangle\otimes\sum_{n=0}^{\infty}c_{n}|n\rangle$. In the rotating framework, we have the effective Hamiltonian illustrated in Fig.~\ref{fig:1}(c) as (see derivations in SM)
\begin{equation}
\begin{aligned} H_\mathrm{eff}=  \sum_{n=0}^{\infty}&\sum_{m=-\infty}^{n}\epsilon m(m+m^{\star})|m,n\rangle\langle m,n|\\
 & +g_{q}\sqrt{n-m}\Big(|m,n\rangle\langle m+1,n|+\mathrm{h.c.}\Big),
\end{aligned}
\end{equation}
where  $m^{\star}$=$-{\hbar(v_{0}\Delta K-\omega_{0})}/{\epsilon}$ is the linear detuning term and $g_{q}$=$-{e\bar{E_{z}}}v_{0}/({2\sqrt{2}\hbar\omega_{0}})$ is the single-photon coupling strength. $e$ is the electron charge, $v_0$ is the initial electron velocity, and $\bar{E_{z}}$ denotes the average longitudinal electric field felt by an electron. Here, the transverse electric field in the coupling region is canceled by the cavity design. The parameter $m^{\star}$ is tunable via the electron initial velocity. For $m^{\star}$=$M$ with $M\! \in \!\mathbb{Z}$, $M$-photon emission process becomes resonant, as shown in Fig.~\ref{fig:1}(b).  $\epsilon$ represents the anharmonicity from nonlinear dispersion relationship of free electrons at low-energy regime~\cite{fan2023JC,eldar2024self}. At large detuning regime, i.e.,, $\epsilon\gg g_{q}\sqrt{\langle n\rangle}$ with average photon number $\langle n\rangle$,  electron populations concentrate on two resonant levels, resulting in quantum Rabi oscillations, as shown in Fig.~\ref{fig:1}(c). 
By compensating momentum mismatch, the periodic structure on the microcavity surface enables large recoil from visible light photons, which is comparable to the kinetic energy of low-energy electrons. While this large recoil is reminiscent of quantum free-electron lasers (FELs)~\cite{Kling2017Quantum}, the amplification mechanism distinguishes both conventional and quantum high-energy FELs~\cite{o2001free,emma2010first,oepts1995free,deacon1977first,Kling2017Quantum}, where the net optical gain depends only on quantum fluctuations rather than photon number $\langle n\rangle$ in our system. See details in SM. This limited net gain prevents FEL systems from achieving exponential gain and lasing without external intervention through micro-bunchings~\cite{Zhang2025superradiance}.
Furthermore, we note that even for $M>1$ cases, this trapping effect also results in an effective Radi oscillation as illustrated in lower panel of Fig.~\ref{fig:1}(c).

Here, we fix interaction time $\tau$ and the injection rate of electron beam $r_\mathrm{e}$.  The photon density matrix satisfies $\mathrm{d}\rho/\mathrm{d}t=(\mathrm{d}\rho/\mathrm{d}t)_\mathrm{gain}+(\mathrm{d}\rho/\mathrm{d}t)_\mathrm{loss}$. The gain density matrix element is given by (see derivations in SM)
\begin{equation}
\begin{aligned}\left(\!\frac{\mathrm{d}\rho_{n,n^{\prime}}}{\mathrm{d}t}\!\!\right)_{\!\!\mathrm{gain}}\!\!= & -r_\mathrm{e}\big[1-\cos(G_{n,M}\tau)\cos(G_{n^{\prime},M}\tau)\big]\rho_{n,n^{\prime}}\\
 & +r_\mathrm{e}\sin(G_{n,M}\tau)\sin(G_{n^{\prime},M}\tau)\rho_{n-M,n^{\prime}-M,}
\end{aligned}
\end{equation}
where $G_{n,M}$=$(\sqrt{n+1}{g_{q}^{M}}/{\epsilon^{M-1}})\prod_{m=1}^{M-1}\frac{\sqrt{n+m+1}}{m(m-M)}$ is the effective coupling strength. 
The dissipation part of  the photon density matrix reads $\left({\mathrm{d}\rho}/{\mathrm{d}t}\right)_{\mathrm{loss}}$=$\kappa n_{\mathrm{th}}\mathcal{D}(a^{\dagger})+\kappa(n_{\mathrm{th}}+1)\mathcal{D}(a)$, where $\kappa$ is the cavity dissipation rate and $n_{\mathrm{th}}$=$\left(e^{{\hbar\omega_{0}}/{k_{b}T}}-1\right)^{-1}$ is the thermal photon number at temperature $T$ with $k_{b}$ the Boltzmann constant. $\mathcal{D}(o)\rho=o\rho o^{\dagger}-\frac{1}{2}\{o^{\dagger}o,\rho\}$ is the Lindblad superoperator and $a$ ($a^{\dagger}$) is the annihilation (creation) operator for optical field. 
In the linear gain regime, the $M$-photon optical gain can be approximated as $\Gamma_{M}$=$r_\mathrm{e}\left({g_{q}^{M}}\tau/{C_{M}\epsilon^{M-1}}\right)^{2}$ with $C_{M}$=$\prod_{m=1}^{M-1}\big[m(M-m)\big]$. $\Gamma_{M}$ decreases with $M$ in the large detunning regime ${g_{q}}/{\epsilon}\ll1$. This allows low-energy free electrons to be a controllable gain medium via velocity-tuned coupling to cavity modes. When the electron-induced gain exceeds the cavity dissipation, lasing is achieved~\cite{haken1984laser,verdeyen1989laser}.

{\it Nonclassical Lasing} --- For $M=1$, we have the steady-state distribution of photon number as
\begin{equation}
P(n)=\rho_{n,n}=P(0)\prod_{j=1}^{n}\frac{\kappa n_{\mathrm{th}}+r_\mathrm{e}\sin^{2}(g_{q}\tau\sqrt{j})/j}{\kappa(n_{\mathrm{th}}+1)},\label{eq:5}
\end{equation}
where $P(0)$ is obtained by the normalization condition $\sum_{n}P(n)$=$1$. In the linear gain regime, the photon number evolves as $\mathrm{d}\langle{n}\rangle/\mathrm{d}t=\big[r_\mathrm{e}(g_{q}\tau)^{2}-\kappa\big]\langle n\rangle+r_\mathrm{e}(g_{q}\tau)^{2}$. Here, when $r_\mathrm{e}\left(g_{q}\tau\right)^{2}>\kappa$, photon number will arise exponentially, which refers to the threshold of the micro-laser being achieved. Thus, the threshold condition is given by
\begin{equation}
\left(g_{q}\tau\right)_{\mathrm{th}}={\langle N_\mathrm{e}\rangle}^{-1/2},
\end{equation}
where $N_\mathrm{e}=r_\mathrm{e}/\kappa$ is the number of electrons that pass through the cavity during the photon lifetime. For a photonic mode with a linewidth of $1\:\mathrm{GHz}$, it is sufficient to have $N_\mathrm{e}=6.25\times10^{3}$, yielding a low threshold $\left(g_{q}\tau\right)_{\mathrm{th}}\approx0.013$, which corresponds to an electron current of $1$~$\mu$A. This contrasts sharply with atomic micro-masers, where preparing more than $10^{9}$ atoms per second in the upper energy state is required---a prohibitive challenge due to the need for optical pumping or Rabi initialization \cite{rempe1987observation,filipowicz1986theory,rempe1990observation,weidinger1999trapping}. 
Differently, our approach utilizes electrons from a conventional gun whose intrinsic energy distribution is inherently suitable for interaction. Even with standard beam conditioning such as focusing and energy filtering, the achievable electron injection rates significantly exceed those of prepared atomic systems, enabling high electron densities.

\begin{figure}[t]
\centering\includegraphics[clip,width=1\columnwidth]{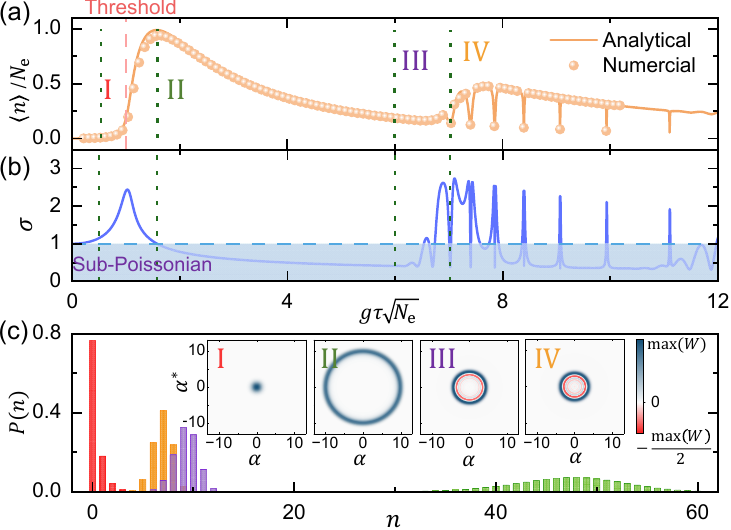}\caption{(a) Analytical (solid curves) and numerical (dots) results of the average photon number $\langle n\rangle$ in an electron-driven laser as a function of $g_{q}\tau$. (b) The normalized variance $\sigma$, quantifying photon number fluctuations, as a function of $g_{q}\tau$. (c) Steady-state photon number distributions. Insets represent  Wigner functions for specific coupling strengths $g_{q}\tau=0.5/\sqrt{N_\mathrm{e}},\:(\pi/2)/\sqrt{N_\mathrm{e}},\:6/\sqrt{N_\mathrm{e}}$ and $\pi/\sqrt{10}$, which are also marked by dotted lines `I--IV' in panels (a) and (b). Here, both numerical and analytical parameters are $v_{0}$=$0.02c$, $\lambda$=$1550$nm, $\kappa$=$2\pi \times1$GHz, $I_\mathrm{e}$=$8$nA, and $\Lambda$=$31$nm.}
\label{fig:2}
\end{figure}

Figures~\ref{fig:2}(a,b) show the average photon number $\langle n\rangle$ and normalized standard deviation $\sigma^{2}$=$\big(\langle n^{2}\rangle-\langle n\rangle^{2}\big)/\langle n\rangle$ as a function of $g_{q}\tau$. Below the threshold value $1/\sqrt{N_\mathrm{e}}$, $\langle n\rangle$ remains negligible with the optical field exhibiting a thermal distribution [Case `I' in Fig.~\ref{fig:2}(c)]. As $g_{q}\tau$ approaches the threshold value, $\langle n\rangle$ increases sharply, reaching approximately $N_\mathrm{e}$ when $g_{q}\tau$=$(\pi/2)/\sqrt{N_\mathrm{e}}$. This regime corresponds to a nearly coherent optical field, driven by electrons undergoing a quarter-period of Rabi oscillation and fully transitioning to the lower energy state [Case `II' in Fig. \ref{fig:2}(c)]. When $g_{q}\tau$ exceeds $(\pi/2)/\sqrt{N_\mathrm{e}}$, photon absorption becomes dominant, reversing the population inversion of electrons and suppressing $\langle n\rangle$. Case `III' in Fig.~\ref{fig:2}(c) illuminates that the photon statistics shift to sub-Poissonian ($\sigma<1$), and the Wigner function exhibits negativity---a signature of nonclassicality~\cite{kenfack2004negativity}. 
Beyond coupling strengths of $g_{q}\tau\approx 2\pi/\sqrt{N_\mathrm{e}}$, $\langle n\rangle$ rises again, but with notably pronounced dips that signify a unique type of Rabi oscillation with a distinct period. 
These oscillations progressively dampen as $g_{q}\tau$ further increases.
Concurrently, photon number manifests a distribution with a few peaks, whose peak values $n_\mathrm{p}<N_\mathrm{e}$ satisfy $\sin^2\!\!\big({g_q\tau\sqrt{n_\mathrm{p}}}\big) \!=\!\kappa n_\mathrm{p}/r_\mathrm{e}$. See details in SM.

By adjusting $g_{q}\tau$, either coherent states or nonclassical states can be generated with thousands of photons in the cavity for injection of electron beams with current intensity below 1$\mu$A. In this stimulated emission approach, acquirement of the coupling strength $g_{q}\tau$ can be further reduced below $0.01$ by improving the cavity quality factor and the electron injecting rate. Compared to single-electron approaches, this one is more experimentally friendly.
The maximum lasing power of this nonclassical microlaser only relates to the injecting rate of electrons and the ratio of external and internal dissipation rates, since each electron emits one photon at most on average for $M=1$. 
For example, the maximum lasing power with a wavelength of $ 1550$ nm is about 1.04$\mu$W, where $I_\mathrm{e}$=1$\mu$A and external dissipation is equal to internal one. Note that a larger $M$ can increase the maximum power by $M$ times, while raising the threshold value accordingly.

\emph{Fock states}---A distinctive regime manifests at discrete coupling strengths $g_{q}\tau$=${j\pi}/{\sqrt{n_j+1}}$, where $\langle n\rangle$ undergoes truncation to an integral value $n_j$, as verified by the pronounced dips in Fig.~\ref{fig:2}(a). At these specific couplings, the emission probability corresponding to the photon number $n_j$ vanishes. This phenomenon is attributable to the interaction of electrons with the cavity field via $2j\pi$-pulse dynamics, which results in their return to their initial state. Such truncation engenders a sub-Poissonian distribution ($\sigma<1$) bounded at $n_j$ [Case `IV' in Fig.~\ref{fig:2}(c)]. Further improvement of the electron injection rate does not lead to a commensurate increase in $\langle n\rangle$; rather, it facilitates the generation of a Fock state $| n_j \rangle$ exhibiting enhanced purity. Of particular note is the suppression of thermal photon excitation---a critical limiting factor in their microwave counterparts. For a set of experimentally feasible parameters $1550$nm cavity mode at $300$K, we have $n_\mathrm{th}$=$4.35\times10^{-14}$, thereby facilitating room-temperature operation with negligible thermal decoherence. 

\begin{figure}[t]
\centering
\includegraphics[clip,width=1\columnwidth]{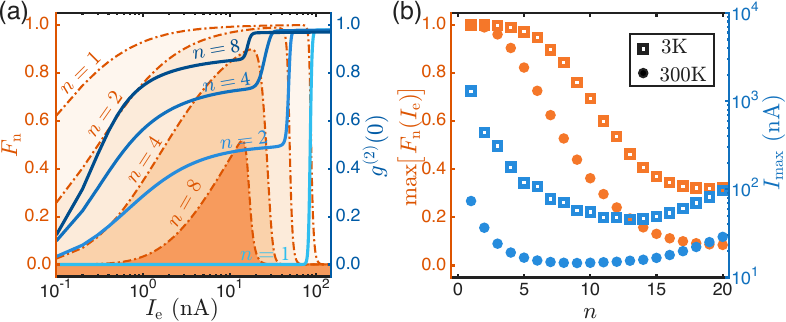}
\caption{(a) Fidelity $F_{n}$ (red dash-dot curves) and two-photon correlation $g^{(2)}$ (blue solid curves) of the Fock state $|n\rangle$ as a function of the current intensity $I_\mathrm{e}$. Numerical parameters are $\lambda$=$1550$nm,  and $g_{q}\tau=\pi/\sqrt{n+1}$ for  target Fock state $|n\rangle$ at temperature $300$K. (b) Maximum fidelity denoted by $\max[F_{n}(I_\mathrm{e})]$ (red dots and squares) and corresponding current requirements $I_{\mathrm{max}}$ (blue dots and squares) for different Fock states at temperatures of $3$K and $300$K, respectively. The parameters are the same as those used in panel (a).}\label{fig:3}
\end{figure}

Here, we use above parameters and finely tune the couplings to match the condition for generating Fock state $|n\rangle$.
Figure~\ref{fig:3}(a) reveals a critical dependence of Fock state fidelity $F_{n}=\sqrt{P(n)}$ for various $I_\mathrm{e}$.  Initially, both the $n$-photon occupation probability and state fidelity increase with $I_\mathrm{e}$ due to the photon number truncation mechanism. However, this ideal trapping condition only holds at zero temperature. At finite temperatures, thermal excitation induces the residual population in the ($n+1$)-photon state. This leads to that the compromised $2\pi$-pulse interaction condition allows photon emission to surpass this designed truncation threshold.  Breakdown of the trapping mechanism also finds quantitative validation in the analytical framework of Eq.~(\ref{eq:5}), where thermal population factor $n_{\mathrm{th}}/(n_{\mathrm{th}}+1)$ can enable gradual population transfer to higher photon states even the sine term at ($n+1$) photons vanishing. 
Consequently, beyond an optimal current threshold, further increase of $I_\mathrm{e}$ causes fidelity to plummet dramatically. 
The second-order correlation $g^{(2)}(0)={\langle a^\dagger a^\dagger aa\rangle}/{\langle a^\dagger a\rangle^2}$ for corresponding Fock states exhibits the purity of single photons. In particular, for state $|1\rangle$, this lasing process can generate ideal single photons for a large regime of $I_\mathrm{e}$.
The trapping efficiency additionally exhibits an inverse relationship with target photon number $n$ that higher Fock states achieve lower maximum fidelities at reduced optimal currents [Fig. \ref{fig:3}(b)]. Importantly, while cavity dissipation rate $\kappa$ influences the required current scale ($I_\mathrm{e}\!\propto\!\kappa$), it preserves the characteristic fidelity-current curve shapes shown in Fig. \ref{fig:3}(a). Thermal suppression strategies are quantitatively compared in Fig. \ref{fig:3}(b). Furthermore, cooling from room temperature ($300$K) to cryogenic conditions ($3$K) remarkably improves the fidelity of $|1\rangle$  with a remarkable increase of corresponding current requirements. Meanwhile, the maximum achievable Fock states with fidelity beyond $90\%$ increase from $|4\rangle$ to $|7\rangle$ under these thermal conditions.

\begin{figure}[t]
\centering
\includegraphics[clip,width=1\columnwidth]{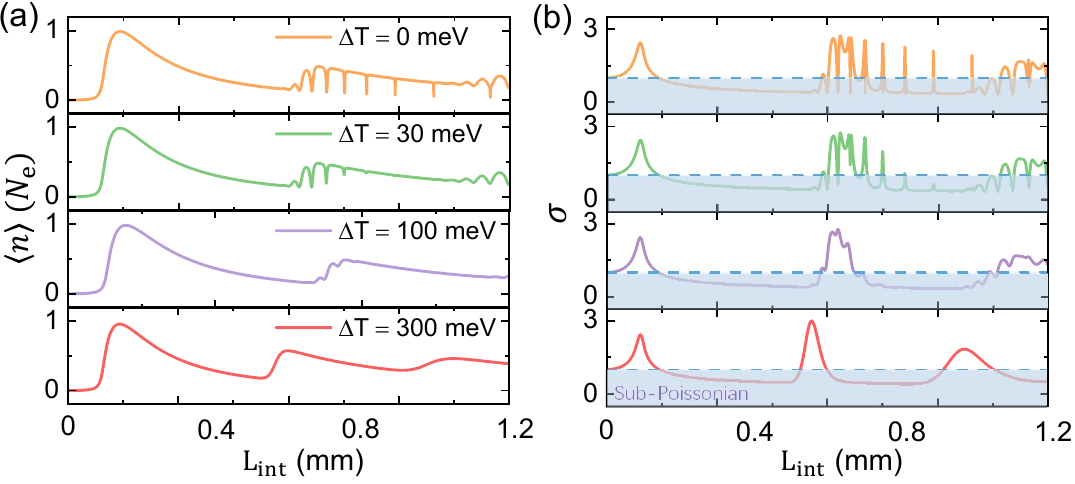}\caption{(a) The average photon number $\langle n\rangle$ as a function of the interaction length $L_{\mathrm{int}}$ under differen energy spreads of the electron beam. Partial parameters are chosen as $T_0=1\,\mathrm{keV}$, $\lambda=1550\,\mathrm{nm}$,  $g_q=30\,\mathrm{GHz}$, $I_e=800\,\mathrm{pA}$, and $\kappa=2\pi\times100\,\mathrm{MHz}$.  (b) Normalized variance $\sigma$ as a function of $g_{q}\tau$ corresponding to different $\Delta T$ in panel (a).}\label{fig:4}
\end{figure}

\emph{Experimental implication}---Undesirable effects, such as injection rate fluctuations, velocity variation, and inter-electron Coulomb interactions, are important to consider during experiments. Temporal variations in electron injection rate $r_\mathrm{e}$ negligibly affect steady-state photon statistics when fluctuation frequencies lie below the electron-photon Rabi frequency, allowing substitution of $r_\mathrm{e}$ with $\langle r_\mathrm{e}\rangle$ in Eq.~(\ref{eq:5}). 
Notably, the energy variation introduces detuning of $\Delta_v\approx\omega_0\Delta T/(2T_0)$, where $T_0$ and $\Delta T$ being the initial kinetic energy and energy spreads  (FWHM of a Gaussian distribution) of electrons, which in turn modifies the photon statistics as
\begin{equation}
{P(n)}\!=\!{P(0)}\!\int \!dv\;G(v)\prod_{j=1}^{n}\frac{\kappa n_{\mathrm{th}}+\frac{4g_q^2r_\mathrm{e}}{\Omega_j^2(v)}\sin^{2}\left(\frac{\Omega_{j}(v)}{2}\tau\right)}{\kappa(n_{\mathrm{th}}+1)},
\end{equation}
where $\Omega_{j}(v)$=$\sqrt{\Delta_v^{2}+4j(g_{q})^{2}}$ is the Rabi frequency for electrons with the velocity $v$ and $G(v)$ is the velocity distribution. Therefore, selecting high values for $T_0$ and $g_q$ (as implemented by $T_0=1\,\mathrm{keV}$ and $g_q=30\,\mathrm{GHz}$) enables robustness to $\Delta T$ in the system. As shown as the average photon number and normalized variance in Fig. \ref{fig:4}, $\Delta T$$>$$100$meV degrade Fock-state trapping at $g_{q}\tau$=$j\pi/\sqrt{n_{j}+1}$, while preserving nonclassical lasing ($\langle n\rangle_\mathrm{max}$$>$$0.9 N_\mathrm{e}$) up to $\Delta T$=$300$meV. The Coulomb interaction between electrons is intrinsically suppressed by the removal of the micro-bunching requirement in high-energy FEL. Additionally, employing cavities with a narrower linewidth $\kappa$ offers a further reduction in the necessary beam current $I_e$ and Coulomb interaction. For instance, with $N_e=50$ fixed, the required $I_e$ is as low as $1.6\,\mathrm{nA}$ and $160\,\mathrm{pA}$ for cavities resonant at $1550\,\mathrm{nm}$ with quality factors of $Q=10^6$ and $Q=10^7$, respectively. For $T_0=1\,\mathrm{keV}$ and $I_e=1.6\,\mathrm{nA}$, average inter-electron distance reaches $1.9$ mm. Neighbor electron generates an electric field $E_{\mathrm{single}}\approx4\times10^{-4}$V/m, which is much weaker than optical-mode fields (kV/m) and could be ignored. 
In summary, we propose generating an electron beam (1nA, 1 keV, 100 meV spread) using a conventional field-emission gun with slight energy selection, then focusing it to a 5 nm radius to interact with a high $Q$ ($Q\sim 10^6$ cavity designed in SM). This scheme is feasible with current technology~\cite{hage2018eV,kociak2014eV,krivanek2009eV,Arend2025Electrons}. 


\emph{Discussion}---In conclusion, this work demonstrates that low-energy free-electron beams, when interacting with a PhC cavity, can reliably generate nonclassical light and high-purity Fock states. When compared to atom-based micromasers, our approach automatically satisfies the initial condition of population as atomic systems by injecting velocity-tailored electrons that fulfill phase-matching criteria, enabling a threshold current of nA level without preparatory operations.  Furthermore, the tunability of low-energy electron energies offers a distinct advantage: the system can produce frequency-agile multiphoton Fock states, where the photon energy is determined by the electron velocity and cavity design rather than fixed atomic transitions, thus overcoming limitations inherent to bound-state systems. 

In addition, this approach offers three fundamental advantages over conventional Fock state preparation schemes. First, the continuous electron beam enables deterministic Fock state generation without post-selection, providing a more straightforward approach than probabilistic preparation methods. Second, the system maintains multi-photon Fock states ($n>1$) through steady-state operation and sequential single-photon emissions, with generation rates governed by the input-output relation:$a_{\mathrm{out}} = \sqrt{\kappa_{\mathrm{ex}}} a$ where $\kappa_{\mathrm{ex}}$ represents the external cavity decay rate. This enables GHz-scale generation rate with rates independent of $n$, contrasting sharply with conventional systems such as quantum dots\cite{Liu2025Quantum,Zhang2025Experimental}, ion trap\cite{Keller2004Continuous}, and single free electron\cite{Theis2024Generation}, where preparation efficiency decreases exponentially with $n$. For the quantum dots system, whose generation rate is highest, the two-photon generation rate is just $28.5\,\mathrm{MHz}$\cite{Liu2025Quantum}, and high-speed generation of higher photon number states is still challenging for experiment. Third, the required quantum nonlinearity emerges intrinsically from the engineered electron-cavity interaction, eliminating the need for additional strong nonlinear materials\cite{Nicholas2023Creating,Rivera2023Nonperturbative} or deep-strong coupling conditions\cite{Nguyen2023Intense} that are challenging to achieve at optical frequencies. The combination of high-speed operation, deterministic production, no strong optical nonlinearity demand, and room-temperature compatibility establishes this free-electron-driven architecture as a promising platform for practical quantum state engineering.

In future studies, extending the framework to $M$-photon gain processes could open pathways to multi-photon cascaded lasing, where electron velocity tuning could synchronize sequential photon emissions for nonlinear quantum electrodynamics studies. Concurrently, introducing of multi-mode cavities may probe nonlinear quantum electrodynamics in unexplored regimes, bridging free-electron physics and quantum optics. These advances establish free-electron-driven microlasers as a scalable platform for quantum technologies. Integrated with photonic circuits, the system could enable room-temperature quantum light sources for sensing, communication, and more---overcoming the cryogenic constraints of atom-based counterparts.

\vspace{0.5cm}

\emph{Acknowledgments}---The work was supported by the National Natural Science Foundation of China (124B2083, 12293052, 12447139, 12375021, and 12247101), the Fundamental Research Funds for the Central Universities, USTC Major Frontier Research Program (LS2030000002). L.Y. acknowledges support by the Zhejiang Provincial Natural Science Foundation of China (LD25A050002). This work was partially carried out at the USTC Center for Micro and Nanoscale Research and Fabrication.

\bibliography{Reference}

\end{document}